\documentclass[12pt]{article}
\usepackage{epsfig,graphics}
\newcommand{\be}{\begin{equation}}
\newcommand{\bea}{\begin{eqnarray}}
\newcommand{\eea}{\end{eqnarray}}
\newcommand{\ba}{\begin{array}}
\newcommand{\ea}{\end{array}}
\newcommand{\ee}{\end{equation}}

\expandafter\ifx\csname mathbbm\endcsname\relax

\else

\fi \textheight 22cm \textwidth 15cm \topmargin 1mm \oddsidemargin
5mm \evensidemargin 5mm

\begin{document}
\begin{titlepage}
\hfill \vbox{
    \halign{#\hfil         \cr
           IPM/P-2005/052 \cr
           hep-th/0511145  \cr
           } 
      }  
\vspace*{20mm}
\begin{center}
{\Large {\bf On Toda Equation and Half BPS Supergravity Solution
in M-Theory}
\\}

\vspace*{15mm} \vspace*{1mm} {Mohammad A. Ganjali}
 \\
\vspace*{1cm}

{\it Institute for Studies in Theoretical Physics
and Mathematics (IPM)\\
 \vspace{3mm}
 Department of Physics, Sharif University of Technology\\
P.O. Box 11365-9161, Tehran, Iran}\\
\vspace*{.5cm} E-mail:Ganjali@theory.ipm.ac.ir\\

 \vspace*{1cm}
\end{center}
\begin{abstract}
Recently, it was shown that half BPS Supergravity solution of
theories with SU(2$|$4) symmetry algebra is given uniformly by
determining a single function which obeys three dimensional
continuous Toda equation. In this paper, we study the scale
invariant solution of Toda equation. Our motivation is that some
solutions of half BPS sector of IIB supergravity, as one excepts
from the fermion description of the theory, are scale invariant.
By defining two auxiliary functions we prove that such solutions
of Toda equation obey cubic algebraic equation. We obtain some
simpl solutions of Toda equation specially, we observe that the
PP-wave solution can be written in this fashion.
\end{abstract}

\end{titlepage}
\section{Introduction}
The quantum theory of space-time has been one of the most
important and basic concepts in physic for long time. String
theory, as a good candidate for theory of every thing, opens
different wonderful windows to the theory of gravity. Holography
\cite{'tHooft:1973jz} is one of the most important and beautiful
subject which is statement about duality between string theory and
gauge theory. AdS/CFT correspondence \cite{Maldacena:1997re} is a
concrete example of such duality which is a strong/weak duality.
It has been found that there is also a weak/weak correspondence
due to large quantum number limit(BMN sector) in gauge theory and
semi-classical strings \cite{Berenstein:2002jq}.

The half BPS sector of such theories has some important property
which helps one to study the AdS/CFT duality in different regime.
In fact, Information coming from BPS sector is protected by
supersymmetry and so all computations can reliable either at weak
coupling or strong coupling.

Recently \cite{Lin:2004nb, Lin:2005nh}, the dual super gravity
solutions of half BPS sector of theories with PSU$(2,2|4)$ SUSY
algebra was found in a uniform way. All such theories are
classified by different subgroup of PSU$(2,2|4)$ with $16$
supercharges which are SU$(2,2|2)$,
PSU$(2|2)\times$PSU$(2|2)\times$ U$(1)$, SU$(2|4)$ and
SuperPoincare part of PSU$(2,2|4)$.

The supergravity solutions are demanded to have some specific
properties. More precisely,

i)solutions should have globally well-defined time-like (or
light-like) Killing vector isometry,

ii)the bosonic part of the isometries should be compact,

iii)solutions should be smooth.

Interestingly, it was shown in \cite{Lin:2004nb,Berenstein:2004kk}
that supergravity theory with PSU$(2|2)\times$PS-\\U$(2|2)\times$
U$(1)$ algebra, which arises in IIB theory naturally, is dual to a
free fermion description of ${\cal N}=4$ super Yang-Mills on
$R\times S^3$. The phase space of two theories is given by two
dimensional space with two topologically different region. The
solution for different boundary condition gives different
deformation of maximally supersymmetric $AdS_5\times S^5$ space.
There have been done a lot of work after that for understanding
these duality better \cite{Liu:2004ru}.

It was also shown that \cite{Lin:2005nh} different gravity
solution with SU$(2|4)$ SUSY algebra can be dual to different
theories including: Plane wave Matrix model(BMN Matrix model),
$2+1$ super Yang-Mills on $R\times S^2$ and ${\cal N}=4$ super
Yang-Mills on $R\times S^3/Z_{k}$. In fact, this solutions which
appear in M-theory are all deformation of $AdS_7\times S^4$ and
$AdS_4\times S^7$ spaces. Finding a free fermion description is an
important aim in completing such duality.

In this paper, at next section, we review the LLM geometry in IIB
supergravity and discuss some basic points about the phase space
of solutions. We observe that some interesting solutions are
invariant under scaling. Then, we introduce the LLM geometry in
M-theory which can be determined by solution of Toda equation. In
section 3 we study the scale invariant solutions of Toda equation
and prove that such solutions obey a cubic algebraic equation. In
section 4 we find some simple scale invariant solutions so called
"separable" solution, PP-wave solution and M-5 brane solution. We
discuss the dual gauge theory of these solutions briefly.
\section{Review of LLM Geometry}
At two next subsection we review briefly the LLM geometry arising
in theories with PSU$(2|2)\times$pSU$(2|2)\times$ U$(1)$ or
SU$(2|4)$ superalgebra. The former case gives the supergravity
solution for half BPS sector in IIB theory and the later, solution
for half BPS sector in M-theory.

\subsection{LLM Geometry in IIB theoty}
For the case PSU$(2|2)\times$PSU$(2|2)\times$ U$(1)$ algebra, the
bosonic part is SO$(4)\times$ SO$(4)\times$ U$(1)$. Considering
the whole requirements about gravity solution, the solution was
found as \cite{Lin:2004nb}
       \bea \label{f1}
            ds^2&=&-h^{-2}(gdt+V_idx^i)^2+h^2(dy^2+dx^idx^i)+ye^Gd\Omega_3^2+
            ye^{-G}d\tilde{\Omega}_3^2,\cr
            h^{-2}&=&y\cosh{G},\;\;\;\;\;\;\;\;y\partial_yV_i=\epsilon_{ij}\partial_jz,\cr
            z&=&\frac{1}{2}\tanh{G}\;\;\;\;\;\;\;\;
       \eea
Here, the dilaton and axion field were assumed to be constant and
the three-form field strengths are zero. The local coordinate $y$
is defined using a closed one from constructed by spinor
bilinears. It has special property since is the product of the
radii of two spheres. Whole solution can be determined by a single
function $z$ which satisfies following differential equation
       \bea \label{f2}
            \partial_i\partial_i z+y\partial_y(\frac{\partial_y
            z}{y})=0.
       \eea
Using the change of variable as $\Phi=z/y^2$, this differential
equation (\ref {f2}) can be reduced to a six dimensional Laplace
equation with spherical symmetry in four of the dimensions, y is
then the radial variable in these four dimensions. At $y=0$ the
product of the radii of two spheres is zero. So one could have
singularities at $y=0$ unless $z$ has a special behavior. As it
was shown in, the smoothness condition implies that we have two
boundary condition on $x_1x_2$ plain.
     \bea \label{f3}
         z=\frac{1}{2}\;\;\;\;S^3\;shrinks,\;\;\;\;\;
         z=-\frac{1}{2}\;\;\;\;\tilde{S}^3\;shrinks
     \eea
So, the moduli space of solution can be determined by specifying
regions on $x_1x_2$ plain that either $z=\frac{1}{2}$ or
$z=-\frac{1}{2}$ (which so called black region or white region
respectively). More interestingly, an arbitrary configuration of
two different regions corresponds to phase space of free fermions
in an specific gauge theory. In fact, the author of has shown that
the half BPS sector of the ${\cal N}=4$ U(N) SYM on $R\times S^3$
is equivalent to an N fermion system in one dimensional harmonic
oscillator potential.

An important observation which has been done in LLM paper is that
the flux of field strengths either for the five form or dual five
form field is proportional to the area of regions where $S^3$ or
$\tilde{S}^3$ shrinks. Using the corresponding fermion phase space
quantization, one can write the precise quantization condition on
the area of the droplets in the $x_1x_2$ plane as
   \bea \label{f4}
        (Area)=4\pi^2l_P^4N\;\;\;\;\;\;\;\;or\;\;\;\;\; \hbar=2\pi
        l_p^4,
   \eea
which means that we have a fundamental length in theory
corresponding to each branes \cite{Lin:2004nb}.

At the level of supergravity solution, such property means that
after a scaling of coordinates $x_i,y$ such that
$x_i\rightarrow\lambda x_i,y\rightarrow\lambda y$ then we have to
have $ds^2\rightarrow\lambda ds^2$
\cite{Lin:2004nb,Horava:2005pv}. Here $\lambda$ is an arbitrary
constant and we actually perform rigid deformation on the shape
of the original configuration. Such scaling behavior is important
because of its relation to Penrose limit \cite{Horava:2005pv}.

Now we focus on finding solutions which are invariant under such
scaling namely
    \bea \label{f5}
         z(\lambda x_i,\lambda y),=z(x_i,y).
    \eea
Such solutions are a particular subset of solutions of (\ref{f2})
,specially when we have multi boundary conditions we can't write
the solution in this way uniformly.

Using the variables $\eta=\frac{x_1}{y}$ and $\zeta=\frac{x_2}{y}$
the equation (\ref{f2}) reduces to
      \bea \label{f6}
        (1+\eta^2)\partial^2_{\eta}z+(1+\zeta^2)\partial^2_{\zeta}z
        +2\eta\zeta\partial^2_{\eta\zeta}z
        +3\eta\partial_{\eta}z+3\zeta\partial_{\zeta}z=0.
      \eea
One can find different kind of solutions of (\ref{f2}) for example
separable solution where $z(\eta,\zeta)=f(\eta)+g(\zeta)$. In this
case one obtains
    \bea   \label{g1}
        f(\eta)&=&k_1+\frac{k'_1}{\sqrt{1+\eta^2}}-\frac{K}{4}\ln{(1+\eta^2)}-\frac{K}{2}
        \int{\frac{\sinh^{-1}\eta}{(1+\eta^2)^{3/2}}d\eta},\cr
        g(\zeta)&=&k_2+\frac{k'_2}{\sqrt{1+\zeta^2}}+\frac{K}{4}\ln{(1+\zeta^2)}+\frac{K}{2}
        \int{\frac{\sinh^{-1}\zeta}{(1+\zeta^2)^{3/2}}d\zeta},
    \eea
where $k_1,k'_1,k_2,k'_2$ and $K$ are arbitrary constants.

For the case where we have an isometry on $x_1$ direction one has
to set $K=0$ and the above differential equation has following
solution
       \bea \label{f7}
            z(\eta)=k_1+k'_2\frac{\eta}{\sqrt{1+\eta^2}},
       \eea
which is the pp-wave solution \cite{Blau:2001ne}. From the
boundary condition (\ref {f3}) we have $k_1=0$ and
$k'_1=\frac{1}{2}$. This pp-wave solution has two different
boundary where extended to infinity. In fact, such extension
allows us to write the solution in terms of $\eta,\zeta$
variables.

Motivated by the above construction of solutions by $\eta,\zeta$
variables, we want to study the supergravity solutions in
M-theory.
\subsection{The LLM Geometry in M-theory}
The other class of subgroup of PSU$(2,2|4)$ with $16$ supercharges
is SU$(2|4)$. The bosonic part of the symmetry is $SO(6)\times
SO(3)\times U(1)$ and the $11$ dimensional supersymmetric
solutions with that symmetry structure is given by
\cite{Lin:2004nb}
     \bea \label{f8}
        ds^2_{11}=&-&4e^{2\lambda}\Bigl[(1+y^2e^{-6\lambda})(dt+V_idx^i)^2+
        \frac{e^{-6\lambda}}{1+y^2e^{-6\lambda}}[dy^2+e^D(dx_1^2+dx_2^2)]\cr
        &+&d\Omega_5^2+y^2e^{-6\lambda}d\tilde{\Omega}_2^2
        \Bigr]\cr
       e^{-6\lambda}&=&\frac{\partial_y
       D}{y(1-y\partial_yD)},\;\;\;\;\;\;\;\;\;\;\;\;\;\;\;\;\;
       V_i=\frac{1}{2}\epsilon_{ij}\partial_jD
     \eea
The function $D$ determines the whole solution and obeys three
dimensional continuous version of the Toda equation
   \bea  \label{f9}
          \partial^2_{x_1}D+\partial^2_{x_2}D+\partial^2_{y}e^{D}=0.
    \eea
The boundary condition for having non singular solution in LLM
construction are that at $y=0$
   \bea   \label{f10}
      \partial_{y}D&=&0,\;\;\;\;D=finite,\;\;\;\;\;S^2\;\;\;\;shrinks\cr
      D&\sim & \ln{y\hspace{3cm}S^5\;\;\;\;shrinks}
  \eea
These conditions ensure that the $y$ coordinate combines with the
sphere coordinates in a non singular fashion.

The moduli space of solutions again is a two dimensional space
with different configuration of black ($S^2$ shrinking sphere) or
white ($S^5$ shrinking sphere) regions.

However, the existence of a free fermion description is not very
clear. In fact, the fluxes of four form field strength and its
dual is given by
   \bea \label{f11}
         N_5\sim\int_{\cal{D}}{dx_1dx_22(y^{-1}e^{D})|_{y=0}},\;\;\;\;\;\;\;
         N_2\sim\int_{\cal{D}}{dx_1dx_22(e^{D})|_{y=0}}.
   \eea
In both cases, the fluxes are given by the area measured with the
metric obtained from $D$. So, one has to find the solution of Toda
equation (\ref {f9}) at first, and then, computes the number of
2-brane or 5- brane. Furthermore, from the nonlinearity of Toda
equation, solutions of such differential equation has not a well
behavior under scaling of coordinates. For example, see the
PP-wave solution (\ref {f27}). But we have an important property
of solution. The three dimensional Toda equation has $SU(\infty)$
$($conformal$)$ symmetry in $x_1x_2$ plane in which the form of
the metric (\ref {f8}) is preserved under such symmetry. Defining
$z=x_1+ix_2$ the symmetry is
       \bea \label{f12}
           z\rightarrow f(z),\;\;\;\;\;\;\;\;\;\;D\rightarrow
           D-\log(|\partial f|^2),
       \eea
where $f(z)$ is an arbitrary holomorphic function of $z$. So, one
may hope that by a conformal transformation obtains a solution
which has well behavior under scaling. At the end, the uniqueness
of the solutions of $nonlinear$ Toda equation with boundary
conditions (\ref{f10}) is another issue which one has to consider
it. For example, for the pp-wave solution (\ref{f28}) in $x_1x_2$
plane we have three nonequal solution for cubic algebraic
equation. Defining
    \bea  \label{g2}
       S_{\pm}=\Bigr(\frac{y^2}{4}+\frac{x^3}{27}\pm
       \sqrt{\frac{y^4}{16}+\frac{y^2x^3}{54}}\Bigr)^{1/3},
    \eea
then equation $(\ref{f28})$ has the following solutions
    \bea   \label{g3}
       (e^D)_1&=&S_++S_-+\frac{x}{3}\cr
       (e^D)_2&=&-\frac{1}{2}(S_++S_-)+\frac{1}{2}i\sqrt{3}(S_+-S_-)+\frac{x}{3}\cr
       (e^D)_3&=&-\frac{1}{2}(S_++S_-)-\frac{1}{2}i\sqrt{3}(S_+-S_-)+\frac{x}{3}.
    \eea
$(e^D)_1$ satisfies the $S^2\;shrinking$ boundary condition for
$x\geq 0$ and $(e^D)_2$ or $(e^D)_3$ (which are nonequal) satisfy
the $S^5\;shrinking$ boundary condition for $x\leq 0$. The flux
driven from $(e^D)_2$ or $(e^D)_3$ are equal up to a minus sign.
Interestingly, in $\rho\theta$ coordinate (which will be defined
in section 4.2), we have a unique solution. Such behavior also
exist for $AdS_4\times S^7$ or $AdS_7\times S^4$ solutions
\cite{Lin:2004nb} where in $(x_1,x_2,y)$ coordinate we have a
quartic algebraic equation.  Considering all above difficulties
cause some ambiguities for having a fermion description at CFT
side.

Motivated by similar case in IIB theory, we will study some
solutions of Toda equation in terms of rational variables. These
solutions, at least, have well behavior under scaling of
coordinates.

Note that if we find a solution $D$ such that $D(\lambda
x_i,\lambda y)=D(x_i,y)$, then at the level of metric (\ref {f8})
we have
    \bea \label{f13}
        (x_i,y)\mapsto (\lambda x_i,\lambda
        y)\Rightarrow\;\;\;\;ds^2\mapsto \lambda^{2/3}ds^2.
    \eea
such behavior comes from the fact that the coordinate $y$ has
dimension $(length)^3$.
\section{Scale Invariant Solution of Toda Equation}
Three dimensional Toda equation is a limit of the exactly solvable
Toda molecule equation and appears in a variety of physical cases,
running from the theory of Hamiltonian systems to general
relativity in the theory of self dual Einstein spaces or in the
problem of finding four dimensional hyper-Kaheler manifold with a
rotational Killing vector \cite{Grassi:1998zt}.

Unfortunately, finding exact solution of the Toda equation is very
hard and only few solutions are known. There are also some methods
based on group theoretical consideration in which one can find the
symmetry structure of the equation and corresponding generator and
then it is possible to reduces the equation to a simpler equation.
The Toda equation allows an $infinite$ dimensional symmetry
algebra, a realization which is given by generators obeying
Virasoro algebra without central charges(Witt algebra). The
reduced equation in this way gives rise to instanton solutions.
For example, one may consider the separable solution in the sense
that $D(x_i,y)=F(x_i)+G(y)$, which cases that the Toda equation
reduces to Liouvile equation which has well known instanton
solution.

For the case that we have an additional isometry on $x_1$
direction one can use the following change of variables
\cite{Ward:1990qt}
    \bea \label{f14}
         e^D=\rho^2,\;\;\;y=\rho\partial_{\rho}V,\;\;\;\;x=\partial_{\theta}V,
    \eea
and reduces the Toda equation to a three dimensional Laplace
equation with cylindrical symmetry as
     \bea \label{f15}
        \frac{1}{\rho}\partial_{\rho}(\rho\partial_{\rho}V)+\partial_{\theta}^2V=0
     \eea
Considering the boundary condition in $\rho\theta$ plane one
realizes that \cite{Lin:2005nh} the problem in this plane reduces
to finding solution for an electroestatic Laplace equation with
some conducting disk located at some constant $\theta_i$ . The
solution is determined by specifying the charge of the disks which
is proportional to $M_2$ brane number and distance between disks
which is proportional to $M_5$ brane number. Even in this case the
exact solution for different configuration of disks is not known
and few solutions were obtained.

In the procedure which we will discuss it, we derive some of
solutions such that have scale invariance property and relate to
solution in electrostatic problem. Interestingly, we will see that
all solutions obtained in this fashion obey a cubic algebraic
equation.

The three dimensional Toda equation $(\ref{f9})$ using change of
variables
    \bea   \label{f16}
             \eta=x_1/y,\;\;\;\;\;\;\;\;\zeta=x_2/y;
    \eea
reduces to
    \bea    \label{f17}
        \partial^2_{\eta}D+\partial^2_{\zeta}D+\eta^2\partial^2_{\eta}e^D
        +\zeta^2\partial^2_{\zeta}e^D
        +2\eta\partial_{\eta}e^D+2\zeta\partial_{\zeta}e^D
        +2\eta\zeta\partial^2_{\eta\zeta}e^D=0.
    \eea
Defining the auxiliary functions $U(\eta,\zeta)$ and
$V(\eta,\zeta)$ $($in which $\partial_{\zeta}U$ and $\partial
_{\eta}V$ are not zero$)$ as
    \bea   \label{f18}
        U(\eta,\zeta)=\partial_{\eta}D+\eta^2\partial_{\eta}e^D,\cr
        V(\eta,\zeta)=\partial_{\zeta}D+\zeta^2\partial_{\zeta}e^D,
    \eea
and using
    \bea  \label{f19}
        \partial^2_{\eta\zeta}e^D&=&e^D(\partial_{\eta}D \partial_{\zeta}D
        +\partial^2_{\eta\zeta}D)=\frac{\partial_{\zeta}U-
        \partial_{\eta}V}{\eta^2-\zeta^2}\cr
        \partial^2_{\eta\zeta}D&=&\frac{1}{2}(\partial_{\zeta}U+\partial_{\eta}V)
        -\frac{1}{2}\frac{\eta^2+\zeta^2}{\eta^2-\zeta^2}
        (\partial_{\zeta}U-\partial_{\eta}V),
    \eea
after some computations one finds
    \bea   \label{f20}
        (e^D)^3+f(\eta,\zeta)(e^D)^2+g(\eta,\zeta)(e^D)+h(\eta,\zeta)=0,
    \eea
where
    \bea \label{f21}
         f(\eta,\zeta)&=&\frac{\eta^2(\eta^2+2\zeta^2)\partial_{\eta}V
         -\zeta^2(\zeta^2+2\eta^2)\partial_{\zeta}U}
         {(\eta^2\zeta^2)(\eta^2\partial_{\eta}V-\zeta^2\partial_{\zeta}U)}\cr
         g(\eta,\zeta)&=&\frac{(\eta^2-\zeta^2)UV+(2\eta^2+\zeta^2)\partial_{\eta}V-
         (2\zeta^2+\eta^2)\partial_{\zeta}U}
         {(\eta^2\zeta^2)(\eta^2\partial_{\eta}V-\zeta^2\partial_{\zeta}U)}\cr
         h(\eta,\zeta)&=&\frac{\partial_{\eta}V-\partial_{\zeta}U}
         {(\eta^2\zeta^2)(\eta^2\partial_{\eta}V-\zeta^2\partial_{\zeta}U)}.
    \eea
 and the Toda equations $(\ref{f9})$ reads
    \bea   \label{f22}
        \partial_{\eta}U+2\frac{\eta\zeta}{\eta^2-\zeta^2}\partial_{\zeta}U(\eta,\zeta)
        +\partial_{\zeta}V-2\frac{\eta\zeta}{\eta^2-\zeta^2}\partial_{\eta}V(\eta,\zeta)
        =0.
    \eea
This is a first order linear partial differential equation. One
can write various solutions for equation $(\ref{f22})$ such
     \bea \label{f23}
         i)\;\;\;\;\;\;U(\eta,\zeta)&=&F_1(\frac{\zeta}{\eta^2+\zeta^2}),\;\;\;\;\;\;
         V(\eta,\zeta)=F_2(\frac{\eta}{\eta^2+\zeta^2})\cr
         ii)\;\;\;\;\;\;U(\eta,\zeta)&=&\frac{b}{a}V(\eta,\zeta)=
         F_3(\frac{a\eta+b\zeta}{\eta^2+\zeta^2}),
     \eea
where $F_i$'s are general functions. Notice that the functions
$F_1$ and $F_2$ are the background solutions for equation
(\ref{f22}). From the linearity of equation (\ref{f22}) the
superposition of any solutions of this equation is also a solution
of $(\ref{f22})$. But, one has to check the consistency condition
in which the obtained solution (\ref {f20}),(\ref {f22}) should
also satisfy equations (\ref {f18}). This consistency check is
hard and we only present some simple solutions using somehow
different idea.

\section{Generating Some solutions}
Solving the Toda equation even in the form (\ref {f18}),(\ref
{f20}) and (\ref {f22}) is very hard and the equation
$(\ref{f20})$ presents an special property of scale invariant
solution. In fact, one has to check the consistency conditions of
solutions. In the following subsections, we present some simple
cases known as separable solution and next to that we find PP-Wave
solution in term of these scale invariant variables. At the next
section, we drive a solution in terms of $\eta$ only by solving
the Toda equation directly.

\subsection{Separable Solution}
The simplest solution can be obtained by rewriting the Toda
equation in terms of ($z,\bar{z},y$) variables and assuming that
$e^D=F(\tilde{\eta})G(\tilde{\zeta})$ where
$\tilde{\eta}=\frac{z}{y-y_0}$ and
$\tilde{\zeta}=\frac{\bar{z}}{y-y_0}$. In this case the solution
is
       \bea \label{f24}
              e^D=(y-y_0)(\frac{\bar{z}^{m+1}}{z^{m}})\;\;\;\;\;\; or
              \;\;\;\;\;\;\; e^D=(y-y_0)(\frac{z^{m}}{\bar{z}^{m+1}})
       \eea
Using conformal transformation (\ref {f12}) and the reality
condition on $D$ one obtains
        \bea \label{f25}
              e^D=c_1y+c_2.
        \eea
This solution doesn't preserve the $S^2 shrinking$ boundary
condition and so is a singular solution.

In electrostatic point of view, we have a configuration with a
line of charge at $\rho=0$ axis in the presence of the external
potential $V_b$ and potential $V$ as \cite{Lin:2005nh}
     \bea \label{f26}
        V=-\frac{\pi
        N}{2k}\log{\rho}+V_b,\;\;\;\;\;\;\;\;V_b=\frac{1}{g_sk}(\rho^2-2\eta^2).
     \eea
After compactifying the $x_1$ direction one obtains a supergravity
solution corresponding to ${\cal N}=4$ super Yang-Mills on
$R\times S^3/Z_k$ \cite{Lin:2005nh,Horowitz:2001uh}. This theory
is an orbifold of ${\cal N}=4$ SYM which the simplest dual
orbifolded supergravity solution is $AdS_5/{Z_k}\times S^5$. The
singularity at $y=0$ corresponds to the $Z_k$ orbifold fixed
points in IIB sense.

As it was shown in, this solution can be viewed as solution for
the near horizon geometry of semilocalised intersecting M2-branes
\cite{Cvetic:2000cj}. It can also be dual to a superconformal
theory, since it is a AdS$_2$ fibration
\cite{Lin:2005nh,Spalinski:2005ha}.
\subsection{PP Wave Solution}
Using changes of variables (\ref {f14}), the Toda equation reduces
to a cylindrically symmetric Laplace equation in three
dimensions(\ref {f15}) which has a polynomial solution as
    \bea \label{f27}
        V=\rho^2\eta-\frac{2}{3}\theta^3
    \eea
The boundary in this case transformed to $\rho=0$ and $\theta=0$.
Solution preserves $S^5$ boundary condition at $\rho=0$ and
$S^2$boundary condition at $\theta=0$. So from electrostatic point
of view, we have an infinite disk at $\theta=0$ and only
$\theta\geq 0$ is physically meaningful.

By the above changes of variables one finds the function $D$ obeys
a cubic algebraic equation as
       \bea \label{f28}
            (e^D)^3-x(e^D)^2-\frac{y^2}{2}=0
       \eea
Now acting a conformal transformation as $z'=(z)^{3/2}$, one can
rewrite the above equation as
      \bea \label{f29}
          (e^{D'})^3-\frac{2}{9}\Bigr((\frac{\tilde{\eta}}{\tilde{\zeta}})^{1/2}+
          (\frac{\tilde{\zeta}}{\tilde{\eta}})^{1/2}\Bigr)(e^{D'})^2
          -\frac{32}{729}\frac{1}{\tilde{\eta}\tilde{\zeta}}=0,
      \eea
where $\tilde{\eta}=\frac{z}{y}$ and
$\tilde{\zeta}=\frac{\bar{z}}{y}$. So we find a solution of (\ref
{f17}) in terms of $\tilde{\eta}\tilde{\zeta}$ variables such that
     \bea \label{f30}
         f=-\frac{2}{9}\Bigr((\frac{\tilde{\eta}}{\tilde{\zeta}})^{1/2}+
          (\frac{\tilde{\zeta}}{\tilde{\eta}})^{1/2}\Bigr),\;\;\;\;\;\;
         g=0,\;\;\;\;\;\;
         h=-\frac{32}{729}\frac{1}{\tilde{\eta}\tilde{\zeta}}.
     \eea
The solution (\ref {f27}),(\ref {f29}) is pp-wave solution in
M-theory with particle by nonzero $-p_{-}$ which are
translationally invariant along $x_{-}$. After compactifying
$x_{-}$ direction one finds dual gravity solution corresponding to
the plane wave (BMN) matrix model
\cite{Berenstein:2002jq,Lin:2004nb}. In fact in IIA variables the
solution in the UV region goes over to the UV region of the
solution for $D_0$ branes
\cite{Lin:2005nh,Maldacena:2002rb,Lin:2004kw}.

For rotationally invariant solutions in $x_1x_2$ plane, one may
perform a conformal transformation to map circular droplet to
strips. In fact, the plane and cylinder can be mapped into each
other conformally. In this case writing two dimensional metric
$dx^2_1+dx^2_2$ in polar coordinates $(r,\theta)$ and using change
of variables as $x_2\rightarrow \ln{r}$ and $D\rightarrow
D+2\ln{x_2}$, one obtains the two dimensional Toda equation which
has a scale invariant solution after performing another conformal
transformation $z'=(z)^{3/2}$.
\subsection{M-5 Brane Solution}
In this section, we consider the case that $D$ is not a function
of $x_1$ and we want to find the scale invariant solution of Toda
equation by solving the differential equation directly. We will
also consider $x_2=x$. Then, using $\eta=\frac{x-x_0}{y-y_0}$ the
Toda equation can be written as
   \bea \label{f31}
       \partial_{\eta}(D)+\eta^2\partial_{\eta}(e^D)=2c,
   \eea
where $c$ is an arbitrary constant. Defining $e^D=\rho^2$ one may
rewrite this equation as
  \bea \label{f32}
       \frac{\partial\eta}{\partial\rho}=\frac{\rho}{c}\eta^2+\frac{1}{\rho
       c}.
  \eea
The solution can be obtained by using the Ricatti change of
variable
    \bea \label{f33}
         \eta(\rho)=-\frac{c}{\rho}\frac{W'(\rho)}{W(\rho)}.
    \eea
Then
    \bea \label{f34}
        W''(\rho)-\frac{1}{\rho}W'(\rho)+\frac{1}{c^2}W(\rho)=0.
    \eea
One can easily find that the solution of this equation has the
following general form,
    \bea \label{f35}
         W(\rho)=c_1\rho J_1(\frac{\rho}{c})+c_2\rho Y_1(\frac{\rho}{c}).
    \eea

where $J_1$ and $Y_1$  are the first and second Bessel functions
respectively. For the case $c_2=0$ using $(\ref{f33})$ we obtain
    \bea \label{f36}
        \eta(\rho)=-
        \frac{1}{\rho}\frac{J_0(\frac{\rho}{c})}{J_1(\frac{\rho}{c})}
    \eea
Although one would like to find an explicit expression for inverse
function $e^D=\rho^2(\eta)$, but we are interested in analyzing
the solution and boundary condition in the $\rho$ coordinate.

For the case where the $S^5$ shrinks one needs as $y\rightarrow
y_0$ then $\eta \sim\frac{1}{\rho^2}$. Since when $\rho
\rightarrow 0$, $J_0(\frac{\rho}{c})\rightarrow 1$ and
$J_1(\frac{\rho}{c})\rightarrow \frac{\rho}{2c}$, the boundary
condition is satisfied.

For the case where $S^2$ shrinks using $\frac{\partial D}{\partial
y}=\frac{2}{\rho}\frac{\partial \rho}{\partial y}$ one finds
      \bea \label{f37}
          \frac{\partial D}{\partial y}=
          \frac{-2c(x-x_0)}{(y-y_0)^2+\rho^2(x-x_0)^2}.
      \eea
We see that for $y_0=0$ the expression(\ref {f37}) will be zero
only at $x=\infty$ and for $y_0\neq 0$ at $x=x_0$ and $x=\infty$.
However, solution(\ref {f36}) shows that $\infty$ corresponds to
$\rho=0$ which means that $D$ is not finite. So the boundary
condition doesn't preserve at $\infty$. For the other case, one
has $\rho\neq 0$ and so the second boundary condition preserved at
$x=x_0$. One may consider (\ref {f36}) as a solution which
produces $AdS_5\times X$ space which $X$ is a six dimensional
compact space. In fact, after an analytic continuation of original
Supergravity solution (\ref{f8}) one finds that this $AdS_5\times
X$ solutions determined with a single function $D$ which obeys the
same Toda equation but it should preserves following boundary
conditions \cite{Lin:2004nb,Maldacena:2000mw}
   \bea   \label{f38}
      \partial_{y}D&=&0,\;\;\;\;\;\;\;\;\;\;D=finite,\;\;\;\;\;S^2\;\;\;\;shrinks\cr
      D&\sim &
      \ln{(y-y_0)\;\;\;\;\;\;y_0\neq0\;\;\;\;\;\;S^5\;\;\;\;shrinks}.
   \eea
One may also consider  $c=0$ at (\ref {f36}), but this does not
imply an interesting solution for the metric.

Let us consider the case $y_0=0$ and $x_0=0$ and $c=\pm 1$. In
fact noticing that if $D(x,y)$ be a solution of Toda equation then
$D(x_i,\lambda y)+2\ln y$ is also a solution of Toda equation,
then, one can generate the solution for the case that $c\neq \pm
1$ using (\ref {f36}).

From the electrostatic point of view, the above solution
corresponds to a potential
$V=J_0(\frac{\rho}{c})e^{\frac{\eta}{c}}$. Both two cases $c=\pm
1$ imply a singular solution but considering the linearity of
Laplace equation one can write a regular solution as
$J_0(\rho)\sinh{\eta}$. Obtaining a regular solution in which the
change of variables $(\ref{f14})$ be well defined, imposes
choosing following solution \cite{Lin:2005nh}
     \bea \label{f39}
         V(\rho,\eta)=I_0(\rho)sin(\eta),
     \eea
where $I_0(\rho)$ is the first modified Bessel function. This can
be done by choosing $c=i$, noticing the fact that only when a
regular solution is at hand one can choose an imaginary value for
$c$. This solution corresponds to two infinite separated disks in
$\rho\theta$ plane.

The solution (\ref{f39}) in IIA language is dual to little string
theory on $R\times S^5$. In large $\rho$ regime the solution
asymptotes to $N$ IIA $NS_5$ branes wrapping on $R\times S^5$
\cite{Berkooz:1997cq}.

\section{Conclusion}
The free fermion description of ${\cal N}=4$ SYM on $R\times S^3$
has had an important rule in AdS/CFT correspondence
\cite{Berenstein:2004kk}. In fact, it was shown that the moduli
space of half BPS sector of IIB supergarivity solutions exactly
mapped to phase space of fermion system
\cite{Lin:2004nb,Lin:2005nh}. This space divided to two regions
where either $S^3$ sphere shrink or $\tilde{S}^3$. Considering the
fact that the area of such regions gives the number of brane and
considering similar description in fermion phase space side, one
realize that the whole solution is invariant under scaling
although, a generic solution may not be scale invariant.

Motivated by such observation, we study some scale invariant
solutions in half BPS sector of supergravity solutions in
M-theory, although the existence of a free fermion description has
not been understood yet. By introducing two auxiliary functions,
we proved that all such solutions can be obtained by solving a
cubic equation.

Unfortunately, even with this simplification, finding the solution
is hard, because that one has to do a consistency check on
solution. We obtain a simple solution so called "separable"
solution which in IIA sense corresponds to solutions of little
string theory. We also write the PP-wave solution in terms of this
homogeneous coordinates. Finally, when we have an addition
isometry, we find the solution of Toda equation directly. Using
this solution one can find a regular solution in electrostatic
point of view.

\section{acknowledgment}
I would like to thank M. Alishahiha, A. Ghodsi, E. Mosaffa and S.
Sheikh-Jabbari for useful discussions and comments.

\end{document}